\documentclass[aps,prl,english,twocolumn,showpacs,preprintnumbers,amsmath,amssymb,floatfix]{revtex4-1}

\usepackage[T1]{fontenc}
\usepackage[latin9]{inputenc}
\usepackage{graphicx}
\usepackage{amssymb}
\usepackage{babel}


\usepackage{babel}
\makeatother

\begin{document}

\title{Casimir interactions in graphene systems}

\author{Bo E. Sernelius}

\affiliation{Division of Theory and Modeling, Department of Physics, 
Chemistry
and Biology, Link\"{o}ping University, SE-581 83 Link\"{o}ping, Sweden}

\email{bos@ifm.liu.se}

\begin{abstract}
The non-retarded Casimir interaction (van der Waals interaction) between two free standing graphene sheets as well as between a graphene sheet and a substrate is determined. An exact analytical expression is given for the dielectric function of graphene along the imaginary frequency axis within the random phase approximation for arbitrary frequency, wave vector, and doping. 
  \end{abstract}

\pacs{73.21.-b, 71.10.-w, 73.22.Lp}

\maketitle
The first reference to the material graphene in the literature was made by Boehm et. al\cite{Boehm} in 1962. With modern technology it is possible to produce large area graphene sheets and graphene has become one of the most advanced two-dimensional (2D) materials of today. It was awarded the nobel prize in physics in 2010. Due to its superior transport properties it has a high potential for technological applications\cite{Novo1, Yuan, Novo2, Berg, Viro}. A free standing graphene sheet has a very interesting band structure. The valence and conduction bands form two sets  of cones. In each set the two cones are aligned above each other with their tips coinciding at the fermi level. Thus, the fermi surface is just two points in the Brillouin zone; the value of the band gap is zero. The energy dispersion in the conduction and valence bands is linear which means that the carriers behave as relativistic particles with zero rest mass. When a graphene layer is formed on a substrate the fermi level moves up or down in energy, the sheet is doped.  A graphene layer interacts with other graphene layers or with a substrate with Casimir\cite{Casi} forces. These forces are very important in graphene systems. Many-body effects also modifies the dispersion of the bands\cite{AAron, BerSer, BSer1, BSer2, BSer3, Shung}. The present work is devoted to the forces.

We begin by calculating the interaction energy between two graphene layers. We limit the calculation to small separations where retardation effects are negligible. Thus we calculate the non-retarded Casimir interaction or in other words the van der Waals interaction. The interaction energy is then nothing but the inter-layer correlation energy\cite{Ser1}.  At zero temperature it is given by 
\begin{equation}
E_c \left( d \right) = \frac{\hbar }{{\left( {2\pi } \right)^2 }}\int\limits_0^\infty  {\int\limits_0^\infty  {d\omega dqq\ln \left\{ {1 - e^{ - 2qd} \left[ {\frac{{\alpha _0 '\left( {q,\omega } \right)}}{{1 + \alpha _0 '\left( {q,\omega } \right)}}} \right]^2 } \right\}} } ,
\label{equ1}
\end{equation}
where $\alpha _0 \left( {q,\omega } \right)$ is the polarizability of one graphene layer. The prime indicates that the function is calculated along the imaginary axis of the complex frequency plane. In terms of the polarizability the dielectric function is given by
\begin{figure}
\includegraphics[width=3.5 cm]{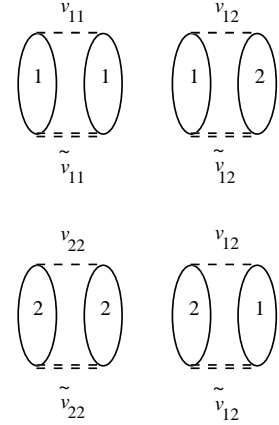}
\caption{Feynman diagrams for the correlation energy in the two graphene sheet system. The ellipses represent polarization bubbles and the dashed lines the interactions indicated in the figure. The numbers 1 and 2 refer to which sheet the electron belongs to. See \cite{Ser1} for details.}
\label{figu1}
\end{figure}
$
\varepsilon \left( {q,\omega } \right) = 1 + \alpha _0 \left( {q,\omega } \right) = 1 - v^{2D} \left( q \right){{\chi \left( {q,\omega } \right)} \mathord{\left/ {\vphantom {{\chi \left( {q,\omega } \right)} \kappa }} \right. \kern-\nulldelimiterspace} \kappa }$, where $v^{2D} \left( q \right) = {{2\pi e^2 } \mathord{\left/ {\vphantom {{2\pi e^2 } q}} \right. \kern-\nulldelimiterspace} q}$ is the two-dimensional fourier transform of the coulomb potential, $\kappa$ is the dielectric constant of the surrounding medium, and ${\chi \left( {q,\omega } \right)}$ the density-density correlation function or polarization bubble. The feynman diagrams representing this energy are given in Fig. \ref{figu1}.
The force is obtained as minus the derivative of the energy with respect to separation, $d$, i.e.,
\begin{equation}
F_c \left( d \right) = \frac{\hbar }{{2\pi ^2 }}\int\limits_0^\infty  {\int\limits_0^\infty  {d\omega dqq^2 \left\{ {1 - e^{ + 2qd} \left[ {\frac{{1 + \alpha _0 '\left( {q,\omega } \right)}}{{\alpha _0 '\left( {q,\omega } \right)}}} \right]^2 } \right\}^{ - 1} } } 
\label{equ2}
\end{equation}

The retarded version of the density-density correlation function,  $\chi \left( {q,\omega } \right)$, for graphene has independently been obtained by Hwang and Das Sarma\cite{Sarma}  and by Wunsch\cite{Wun} et al.  Here we present our general expression in the complex frequency plane, away from the real axis. In particular we give the result on the imaginary axis, which is where we perform the present calculations. We, just as in \cite{Sarma,Wun}, include the contribution from the conduction and valence bands,  only, and assume that the linear dispersion of the bands extends forever. In the real system the dispersion starts to deviate from linear at some point and transitions from the occupied core states to empty bands higher up in energy will contribute to some extent to the screening. We neglect this here. Let us first begin with an undoped graphene sheet. In a general point in the complex frequency plane, away from the real axis the density-density correlation function is
\begin{equation}
\chi \left( {{\bf{q}},z} \right) =  - \frac{g}{{16\hbar }}\frac{{q^2 }}{{\sqrt {v^2 q^2  - z^2 } }},
\label{equ3}
\end{equation}
where $v$ is the carrier velocity which is a constant in grapene ($E =  \pm \hbar vk$), and $g$ represents the degeneracy parameter with the value of 4 (a factor of 2 for spin and a factor of 2 for the cone degeneracy.)

With this particular screening it turns out that $\alpha _0 '\left( {{{\bf{q}} \mathord{\left/ {\vphantom {{\bf{q}} d}} \right.
 \kern-\nulldelimiterspace} d},{\omega  \mathord{\left/ {\vphantom {\omega  d}} \right. \kern-\nulldelimiterspace} d}} \right) = \alpha _0 '\left( {{\bf{q}},\omega } \right)$ and the separation dependence of the interaction becomes very simple, after a change of dummy variables,
\begin{equation}
\begin{array}{*{20}c}
   {E_c \left( d \right) = \frac{1}{{d^3 }}\frac{\hbar }{{\left( {2\pi } \right)^2 }}\int\limits_0^\infty  {\int\limits_0^\infty  {d\omega dqq\ln \left\{ {1 - e^{ - 2q} \left[ {\frac{{\alpha _0' \left( {q,\omega } \right)}}{{1 + \alpha _0' \left( {q,\omega } \right)}}} \right]^2 } \right\}} } }  \\
   {\quad \quad \quad  \approx 2.156\frac{1}{{d^3 }}{N \mathord{\left/
 {\vphantom {N m}} \right.
 \kern-\nulldelimiterspace} m},}  \\
\end{array}
 \label{equ4}
\end{equation}
with the value for $v$ chosen\cite{Wun} to be $8.73723 \times 10^5 {m \mathord{\left/ {\vphantom {m s}} \right. \kern-\nulldelimiterspace} s}$; $d$ is the distance in \AA.

When the graphene sheet is doped the expression for the density-density correlation function is much more complicated. In the two next equations we use dimension-less variables. The variable  $x$ is the momentum divided by $2k_F $; the variable $y$ is the frequency divided by ${{2E_F } \mathord{\left/ {\vphantom {{2E_F } \hbar }} \right. \kern-\nulldelimiterspace} \hbar } $; the variable  $z$ is a general complex valued frequency divided by ${{2E_F } \mathord{\left/ {\vphantom {{2E_F } \hbar }} \right. \kern-\nulldelimiterspace} \hbar } $.  

In a general point in the complex frequency plane, away from the real axis the density-density correlation function is

\begin{widetext}
\begin{equation}
\chi \left( {{\bf{q}},z} \right) =  - D_0 \left\{ {1 + \frac{{x^2 }}{{4\sqrt {x^2  - z^2 } }}\left[ {\pi  - {\rm{asin}}\left( {\frac{{1 - z}}{x}} \right) - {\rm{asin}}\left( {\frac{{1 + z}}{x}} \right) + \frac{{z - 1}}{x}\sqrt {1 - \left( {\frac{{z - 1}}{x}} \right)^2 }  - \frac{{z + 1}}{x}\sqrt {1 - \left( {\frac{{z + 1}}{x}} \right)^2 } } \right]} \right\}.
\label{equ5}
\end{equation}
On the imaginary axis it is 
\begin{equation}
\begin{array}{l}
 \chi '\left( {{\bf{q}},\omega } \right)
  = \chi \left( {{\bf{q}},i\omega } \right) =  - D_0 \left\{ {1 + \frac{{x^2 }}{{4\sqrt {y^2  + x^2 } }}\left[ {\pi  - {\rm{atan}}\left[ {\frac{{2\left\{ {\left[ {x^2 \left( {y^2  - 1} \right) + \left( {y^2  + 1} \right)^2 } \right]^2  + \left( {2yx^2 } \right)^2 } \right\}^{{1 \mathord{\left/
 {\vphantom {1 4}} \right.
 \kern-\nulldelimiterspace} 4}} \sin \left\{ {\frac{1}{2}{\rm{atan}}\left[ {\frac{{2yx^2 }}{{x^2 \left( {y^2  - 1} \right) + \left( {y^2  + 1} \right)^2 }}} \right]} \right\}}}{{\sqrt {\left( {x^2  + y^2  - 1} \right)^2  + \left( {2y} \right)^2 }  - \left( {y^2  + 1} \right)}}} \right]} \right.} \right. \\ 
 \left. {\left. {\quad \quad \quad \quad \quad \quad \quad  \quad \quad  \quad  \quad \quad \quad \quad \quad \quad \quad  - \frac{{\sqrt { - 2x^2 \left( {y^2  - 1} \right) - 2\left( {y^4  - 6y^2  + 1} \right) + 2\left( {y^2  + 1} \right)\sqrt {x^4  + 2x^2 \left( {y^2  - 1} \right) + \left( {y^2  + 1} \right)^2 } } }}{{x^2 }}} \right]} \right\};\quad 0 \le {\rm{atan}} < \pi,  \\ 
 \end{array}
 \label{equ6}
\end{equation}

\end{widetext}
where the density of states at the fermi level, $D_0 $, is
\begin{equation}
D_0  = \frac{{gE_F }}{{2\pi \left( {\hbar v} \right)^2 }} = \frac{{gk_F^2 }}{{2\pi E_F }} = \frac{{2n}}{{E_F }} = \sqrt {\frac{{gn}}{{\pi v^2 }}}. 
\label{equ7}
\end{equation}
Here $n$ is the doping concentration. The same result holds for excess of electrons and excess of holes.

\begin{figure}
\includegraphics[width=7cm]{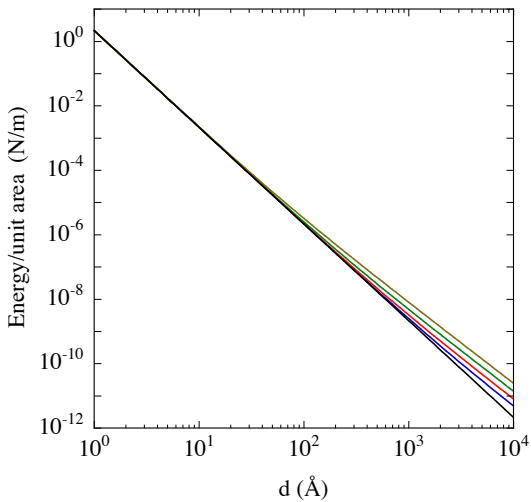}
\caption{The attractive interaction energy between two graphene sheets. The straight line is for undoped sheets, while the bent curves are for doping densities $1 \times 10^{10}, {\rm{ }}1 \times 10^{11}, {\rm{ }}1 \times 10^{12}, {\rm{ and }} {\kern 1pt} 1 \times 10^{13} {\rm{ cm }}^{-2} $, respectively, counted from below. }
\label{figu2}
\end{figure}

\begin{figure}
\includegraphics[width=7cm]{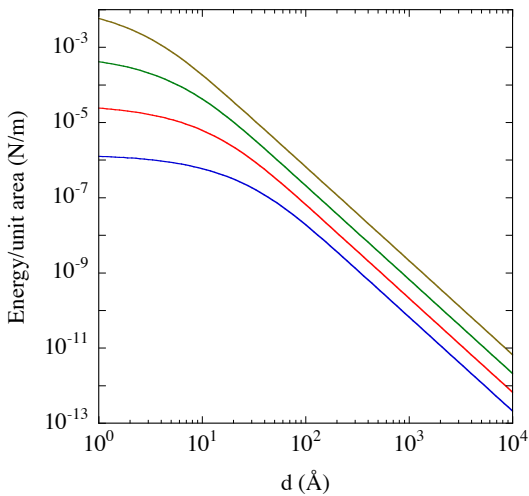}
\caption{The attractive interaction energy between two 2D metallic layers. The curves are for doping densities $1 \times 10^{10}, {\rm{ }}1 \times 10^{11}, {\rm{ }}1 \times 10^{12}, {\rm{ and }} {\kern 1pt} 1 \times 10^{13} {\rm{ cm }}^{-2} $, respectively, counted from below.}
\label{figu3}
\end{figure}

The numerical results for the size of the interaction energy between two undoped graphene layers in vacuum is shown as the straight line in Fig. \ref{figu2}. The bent curves are valid for doping concentrations  $1 \times 10^{10} ,{\rm{ }}1 \times 10^{11} ,{\rm{ }}1 \times 10^{12} , {\rm{ and }}{\kern 1pt}  1 \times 10^{13} {\rm{ cm }}^{-2} $, respectively, counted from below. The interaction energy is negative, leading to an attractive force.
The interaction between two 2D metallic sheets was derived in \cite{Ser1}. To illustrate the difference between the two systems we show in Fig. \ref{figu3} the corresponding results for two 2D metallic sheets in vacuum for the same set of carrier concentrations; we have used the effective mass unity for the carriers.
The van der Waals interaction between two atoms varies with separation, $d$, as $d^{ - 6}$. Between two half spaces it varies as $d^{ - 2}$. We found in \cite{Ser1} that the system of two 2D metallic sheets has a separation dependence with a fractional power, v.i.z., $d^{ - {5 \mathord{\left/ {\vphantom {5 2}} \right. \kern-\nulldelimiterspace} 2}}$ (see the straight part of the curves in Fig. \ref{figu3}.) Here, we have found a system with yet another separation dependence. For two undoped graphene layers the van der Waals interaction varies as $d^{ - 3}$. This is the same dependence as for the Casimir interaction between two half spaces. The origin of the half-integer behavior for the 2D metal sheets is the square root behavior of the dispersion curve for 2D plasmons. In undoped graphene the plasmon dispersion curve is linear. In the doped case, however, the plasmon dispersion curve attains the square root behavior and for larger separations the interaction varies as $d^{ - {5 \mathord{\left/ {\vphantom {5 2}} \right. \kern-\nulldelimiterspace} 2}}$ (see the rightmost part of Fig. \ref{figu2}). For small enough separations the contribution to the interaction from the free carriers saturates. This is obvious in both Figs. \ref{figu2} and \ref{figu3}.

Before we treat next geometry which is a graphene sheet above a substrate it is illustrating to rederive Eq. (\ref{equ1}) in a different way. In \cite{Ser1} we derived it in two alternative ways; here we do it in yet another way. Let us assume that we have an induced carrier distribution, $\rho_1 \left( {{\bf{q}},\omega } \right)$, in sheet 1. This gives rise to the potential $
v\left( {{\bf{q}},\omega } \right) =  v^{2D} \left( q \right)\rho _1 \left( {{\bf{q}},\omega } \right)
$ in sheet 1. It results in the potential $
\exp \left( { - qd} \right)v^{2D} \left( q \right)\rho _1 \left( {{\bf{q}},\omega } \right)
$ in sheet 2, which is parallel with sheet 1, the distance $d$ away. This potential is screened by the carriers in sheet 2. The resulting potential is $
{{\exp \left( { - qd} \right)v^{2D} \left( q \right)\rho _1 \left( {{\bf{q}},\omega } \right)} \mathord{\left/
 {\vphantom {{\exp \left( { - qd} \right)v^{2D} \left( q \right)\rho _1 \left( {{\bf{q}},\omega } \right)} {\left[ {1 + \alpha _0 \left( {{\bf{q}},\omega } \right)} \right]}}} \right.
 \kern-\nulldelimiterspace} {\left[ {1 + \alpha _0 \left( {{\bf{q}},\omega } \right)} \right]}}$. This gives rise to an induced carrier distribution in sheet 2, 
\begin{equation}
\rho _2 \left( {{\bf{q}},\omega } \right) = \chi \left( {{\bf{q}},\omega } \right)e^{ - qd} v^{2D} \left( q \right)\frac{{\rho _1 \left( {{\bf{q}},\omega } \right)}}{{1 + \alpha _0 \left( {{\bf{q}},\omega } \right)}}.
\label{equ8}
\end{equation}
In complete analogy, this carrier distribution in sheet 2 gives rise to a carrier distribution in sheet 1,
\begin{equation}
\rho _1 \left( {{\bf{q}},\omega } \right) = \chi \left( {{\bf{q}},\omega } \right)e^{ - qd} v^{2D} \left( q \right)\frac{{\rho _2 \left( {{\bf{q}},\omega } \right)}}{{1 + \alpha _0 \left( {{\bf{q}},\omega } \right)}}.
\label{equ9}
\end{equation}
To find the condition for self-sustained fields, normal modes, we let this induced carrier density in sheet number 1 be the carrier density we started from. This leads to
\begin{equation}
1 - e^{ - 2qd} \left[ {\frac{{\alpha _0 \left( {{\bf{q}},\omega } \right)}}{{1 + \alpha _0 \left( {{\bf{q}},\omega } \right)}}} \right]^2  = 0.
\label{equ10}
\end{equation}
The left hand side of this equation is exactly the argument of the logarithm in Eq. (\ref{equ1}). This equation was derived using many-body theory. In \cite{Ser2} the interaction energy is derived from the electromagnetic normal modes of the system. One ends up with an identical expression to the one in  Eq. (\ref{equ1}) where now the argument of the logarithm is the function in the condition for normal modes. If we have a 2D layer (like a graphene sheet) the distance $d$ above a perfect metal substrate the procedure is very similar. We start with an induced mirror carrier density,  $\rho_1 \left( {{\bf{q}},\omega } \right)$, in the substrate. The induced carrier density in the graphene sheet is given by the expression in Eq. (\ref{equ8}) except that now the distance beween the mirror charge and the graphene layer is $2d$ instead of $d$. Eq. (\ref{equ9}) is then replaced by $\rho _1 \left( {{\bf{q}},\omega } \right) =  - \rho _2 \left( {{\bf{q}},\omega } \right)$, according to the result for a mirror charge at a perfect metal interface. The condition for modes becomes
\begin{equation}
1 - e^{ - 2qd} \frac{{\alpha _0 \left( {{\bf{q}},\omega } \right)}}{{1 + \alpha _0 \left( {{\bf{q}},\omega } \right)}} = 0,
\label{equ11}
\end{equation}
and the interaction energy is
\begin{equation}
E_c \left( d \right) = \frac{\hbar }{{\left( {2\pi } \right)^2 }}\int\limits_0^\infty  {\int\limits_0^\infty  {d\omega dqq\ln \left\{ {1 - e^{ - 2qd} \left[ {\frac{{\alpha _0 '\left( {q,\omega } \right)}}{{1 + \alpha _0 '\left( {q,\omega } \right)}}} \right]} \right\}} } .
\label{equ12}
\end{equation}
Unfortunately, for a graphene layer above a perfect metal substrate the integral does not converge, without  a cutoff; the polarizability dies off too slowly with frequency; for a 2D metallic sheet, on the other hand, above a perfect metal substrate there is no problem. 

\begin{figure}
\includegraphics[width=7cm]{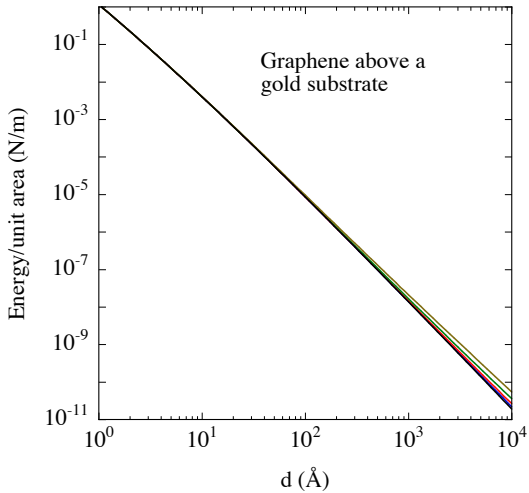}
\caption{The attractive interaction energy between a graphene sheet and a gold substrate. The lowest curve is for an undoped sheet, while the other curves are for doping densities $1 \times 10^{10}, {\rm{ }}1 \times 10^{11}, {\rm{ }}1 \times 10^{12}, {\rm{ and }} {\kern 1pt} 1 \times 10^{13} {\rm{ cm }}^{-2} $, respectively, counted from below.}
\label{figu4}
\end{figure}

Next we focus on a graphene layer above a real substrate. The derivation is the same as for a perfect metal substrate until the last step. Now, using the theory of image charges we realize that the relation between the induced carrier densities is
\begin{equation}
\rho _1 \left( {{\bf{q}},\omega } \right) =  - \rho _2 \left( {{\bf{q}},\omega } \right)\frac{{\varepsilon _s \left( \omega  \right) - 1}}{{\varepsilon _s \left( \omega  \right) + 1}},
\label{equ13}
\end{equation}
and the condition for normal modes becomes
\begin{equation}
1 - e^{ - 2qd} \frac{{\alpha _0 \left( {{\bf{q}},\omega } \right)}}{{1 + \alpha _0 \left( {{\bf{q}},\omega } \right)}}\frac{{\varepsilon _s \left( \omega  \right) - 1}}{{\varepsilon _s \left( \omega  \right) + 1}} = 0.
\label{equ14}
\end{equation}
We have neglected spatial dispersion in the substrate; inclusion of spatial dispersion would lead to a higher order of complexity\cite{Ser3} and would have negligible effects on the present results.
From the condition in Eq. (\ref{equ14}) follows that the interaction energy is given by
 \begin{widetext}
\begin{equation}
E_c \left( d \right) = \frac{\hbar }{{\left( {2\pi } \right)^2 }}\int\limits_0^\infty  {\int\limits_0^\infty  {d\omega dqq\ln \left\{ {1 - e^{ - 2qd} \left[ {\frac{{\alpha _0 '\left( {q,\omega } \right)}}{{1 + \alpha _0 '\left( {q,\omega } \right)}}\frac{{\varepsilon _s '\left( \omega  \right) - 1}}{{\varepsilon _s '\left( \omega  \right) + 1}}} \right]} \right\}} }. 
\label{equ15}
\end{equation}
 \end{widetext}
The result for a graphene sheet above a gold substrate is shown in Fig. \ref{figu4}. The dielectric function of gold along the imaginary frequency axis was obtained from experimental data extrapolated in a way described in \cite{BosSer} and using a modified Kramers Kronig dispersion relation (see Eq. (6.75) in \cite{Ser2}.)

In summary, we have derived and calculated the non-retarded Casimir interaction between two graphene sheets and between a graphene sheet and a substrate. We found the interaction energy between two virgin graphene sheets varies with separation, $d$, as $d^{ - 3} $ (the force as $d^{ - 4} $.) This is an unexpected result. For a pair of films summation over pair interactions leads to a $d^{ - 4} $ dependence. The interaction between doped graphene sheets has a more complex separation dependence. We have derived the dielectric function of graphene along the imaginary frequency axis within the random phase approximation for arbitrary frequency, wave vector, and doping. These results are needed for the present calculations and for future calculations of many-body effects in graphene.

\begin{acknowledgments}
The research was sponsored by the
VR-contract No:70529001 and support from the VR Linn\'{e} Centre LiLi-NFM
and from CTS is gratefully acknowledged.
\end{acknowledgments}

\end{document}